# Morphological Transformations of Diblock Copolymers in Binary Solvents: A Simulation Study

*Zheng Wang, Yuhua Yin, Run Jiang, Baohui Li* *

School of Physics, Key Laboratory of Functional Polymer Materials of Ministry of Education, Nankai University, and Collaborative Innovation Center of Chemical Science and Engineering (Tianjin), Tianjin, 300071, China

---

* To whom correspondence should be addressed.

* E-mail: baohui@nankai.edu.cn (B.L.).

# ABSTRACT

Morphological transformations of amphiphilic AB diblock copolymers in mixtures of a common solvent (S1) and a selective solvent (S2) for the B block are studied using the simulated annealing method. We focus on the morphological transformation depending on the fraction of the selective solvent $C_{S2}$, the concentration of the polymer $C_P$, and the polymer–solvent interactions $\varepsilon_{ij}$ ($i$ = A, B; $j$ = S1, S2). Morphology diagrams are constructed as functions of $C_P$, $C_{S2}$, and/or $\varepsilon_{AS2}$. The copolymer morphological sequence from dissolved → sphere → rod → ring/cage → vesicle is obtained upon increasing $C_{S2}$ at a fixed $C_P$. This morphology sequence is consistent with previous experimental observations. It is found that the selectivity of the selective solvent affects the self-assembled microstructure significantly. In particular, when the interaction $\varepsilon_{BS2}$ is negative, aggregates of stacked lamellae dominate the diagram. The mechanisms of aggregate transformation and the formation of stacked lamellar aggregates are discussed by analyzing variations of the average contact numbers of the A or B monomers with monomers and with molecules of the two types of solvent, as well as the mean square end-to-end distances of chains. It is found that the basic morphological sequence of spheres to rods to vesicles and the stacked lamellar aggregates result from competition between the interfacial energy and the chain conformational entropy. Analysis of the vesicle structure reveals that the vesicle size increases with increasing $C_P$ or with decreasing $C_{S2}$, but remains almost unchanged with variations in $\varepsilon_{AS2}$.

# I. INTRODUCTION

Amphiphilic AB diblock copolymers dissolved in a selective solvent, which is good for one block but poor for the other one, can spontaneously self-assemble into aggregates of various morphologies, such as spheres, rods, rings, and vesicles. Because of the important biological insights gained from these aggregates and their potential biomedical applications as microcapsules[1] or cell membrane mimetics[2-5] and for drug delivery or targeting release,[6-8] copolymer solutions have attracted much scientific interest.[9-20]

Experimentally, the cosolvent method is frequently used to prepare aggregates in solution. In brief, amphiphilic copolymers are dissolved first in a common solvent, such as *N,N'*-Dimethylformamide (DMF), dioxane, or tetrahydrofuran (THF), in which the copolymer chains are dispersed. Subsequently, a selective solvent, such as water, is slowly added to the solution until the water content (generally 25–50 wt%) is much higher than that at which aggregation starts. Then, the aggregates are usually quenched in excess water to freeze the kinetic processes and morphologies. Finally, the common solvent is removed by dialysis of the resulting solution against water.[21] It has also been found that the copolymer composition and concentration, the selective solvent fraction, and the nature of the common solvent are all important factors that may affect the aggregate morphology significantly. Many studies have been carried out to examine these factors.[21-24] A wide range of crew-cut aggregates of different morphologies has been prepared from dilute solutions of asymmetric amphiphilic diblock copolymers. These aggregates include spherical micelles, rods, bicontinuous structures, lamellae, vesicles, large compound micelles, large compound vesicles, and tubules. Overall, more than 20 morphologies and various morphological transitions have been identified by Eisenberg's group.[21,22] Du et al. examined morphology transformations by varying the chain length and the selective solvent fraction.[23] In their study, various vesicular morphologies, such as entrapped vesicles, hollow concentric vesicles, ellipsoidal vesicles, and open bending lamellae, were found to coexist for poly(DLlactide)-*b*-poly(ethyleneglycol) (PLA$_{212}$-PEG$_{44}$) in THF/H$_2$O and dioxane/H$_2$O mixed solvents with 30 and 40 wt% water contents.[23] Huang et al. obtained donut-shaped toroidal micelles of highly uniform shape and size, formed by polyisoprene-*b*-poly(2-vinylpyridine) (PI$_{1100}$-b-P2VP$_{220}$) in THF/ethanol solvent mixtures.[24] Denkova et al. investigated the morphologies and sizes of micellar aggregates formed by the triblock copolymer P123 (EO$_{20}$PO$_{70}$EO$_{20}$) in a mixture of DMF, as an aprotic solvent, and water. They found that bicontinuous micelles with distinct patterns formed at water contents between about 27 and 35 wt%, in coexistence with very long, non-branched, worm-like micelles.[25] In the study of Zhonget al., triblock copolymer poly(acrylic acid)-*b*-poly(methyl acrylate)-*b*-polystyrene (PAA-b-PMA-b-PS) self-assembled into a single or double helix consisting of cylindrical micelles by changing the ratio of water and THF.[26] Han et al. reported that with the addition of water, cylindrical micelles formed by triblock copolymer poly(tert-butyl acrylate)-*b*-poly(2-cinnamoyloxyethyl methacrylate)-*b*-poly(N, Ndimethyl methacrylate) (PtBA-b-PCEMA-b-PDMAEMA) in methanol transformed into a fused horseshoe phase.[27] McKenzie et al. showed that nanospheres with internal bicontinuous morphologies could be obtained from simple, amorphous diblock copolymer poly(ethylene oxide)-*b*-poly(n-butyl methacrylate) (PEO-b-PBMA) in a cosolvent of THF/water.[28] It is found that the aggregates always start from spherical micelles when a selective solvent such as water is added, regardless of the nature of the common solventin experiments.

In addition to extensive experimental investigations, theoretical studies and computer simulations provide powerful tools for studying the self-assembly of block copolymer solutions.[29-35]

The majority of investigations have focused on the effect of the properties of the copolymer or selective solvent on aggregate morphologies, such as copolymer composition, volume fraction, configuration, and selectivity strength. Using simulated annealing, Sun et al. studied the self-assembly of diblock copolymers in selective solvents (using only one type of solvent). The simulation results illustrated that the self-assembled morphologies of the copolymer aggregates strongly depend on the interactions between the core-forming blocks and the solvents, as well as the length of the corona-forming blocks. A transition sequence of disordered state to spherical micelles to short rod-like micelles to long rod-like micelles to onion-like aggregates was observed for copolymers as the core–solvent interactions increased or the length of the corona-forming blocks decreased.[29] Using self-consistent field theory, Wang et al. investigated the self-assembly of amphiphilic AB diblock copolymers with different molecular sizes of solvent in solution.[30] They constructed a phase diagram by continuously varying the solvent size and the polymer concentration, and found that the aggregate concentration of the amphiphilic AB diblock copolymer decreases in solution with larger solvent sizes. Using Monte Carlo simulations, Jiang's group found that the hydrophilicity of blocks A and C was the key parameter for vesicle formation and the microphase behavior of ABC triblock copolymers in selective solvents for A and C blocks.[31] For an A-B-C-A tetrablock copolymer in selective solvents, the chain length ratio and hydrophobicity of blocks B and C were key factors in determining the hydrophobic layer structure of the vesicles.[32] In their study on asymmetric vesicles constructed from an AB/CB diblock copolymer mixture in a selective solvent for A and C blocks, they found that the vesicle structure sequence depends on the composition of the mixture, the chain length of the hydrophilic block, and the solution pH.[33] Liang's group investigated the microstructures assembled from amphiphilic triblock copolymers in selective solvents using the DPD approach, and reported the different pathways involved in the formation of aggregates.[34] Kong et al. studied the self-assembly of ABC star terpolymers composed of a solvophilic A arm and two solvophobic B and C arms in selective solvents. Multicompartment micelles have been predicted, and the results revealed that the overall micelle morphology is largely controlled by the volume fraction of the solvophilic A arms, whereas the internal compartmented and/or segregated structures depend on the ratio between the volume fractions of the two solvophobic arms.[35] Although there are many theoretical and simulation studies on the self-assembly of block copolymer in dilute solution, usually only one type of solvent has been considered. Results from such studies are quantitatively in agreement with the corresponding experimental results, however, systems in which just one type of solvent is used are fundamentally different from those in which two types of solvents are used.

For self-assembly of diblock copolymers in solution using two types of solvents, a common solvent and a selective solvent, we are aware of only one report. Using self-consistent field theory, Huang and Hsu investigated the effects of solvent immiscibility on the phase behavior and microstructural length scales of a diblock copolymer in the presence of two solvents: a neutral solvent and a slightly B-selective solvent.[36] They found that the ordered microphase region in concentrated solutions enlarges as the immiscibility increases and an ordered structure forms that can undergo macrophase separation. In the present work, we report a systematic study of the self-assembly of AB diblock copolymers in binary solvents consisting of a common solvent and a selective solvent using the simulated annealing technique. Unlike the concentrated solutions considered in Ref. 36, dilute solutions are investigated here. - A series of phase diagrams are constructed as functions of various parameters, including the selective solvent fraction, polymer concentration, and polymer–solvent interactions. Rich copolymer morphological transformations are obtained and compared with those previously observed experimentally. It is found that the selectivity of the selective solvent affects the self-assembled microstructure significantly, and aggregates of stacked lamellae dominate the diagrams under certain conditions. The mechanisms of aggregate transformation and the formation of stacked lamellar aggregates are discussed by analyzing variations in the average contact numbers between the A or B monomers and the two types of solvent molecules, as well as the mean square end-to-end distance of the copolymer chains. In addition, the vesicle structure is analyzed.

## II. MODEL AND METHOD

In the current study, the self-assembly of AB diblock copolymers in a binary mixture of a common and a selective solvents was investigated using the simulated annealing method applied to the "single-site bond fluctuation" model of polymers.[37,38] Previous studies on this model system established that this approach provides an efficient methodology for studying the self-assembly of block copolymers in solution.[29] For completeness, the model and algorithm are reviewed briefly below, and a detailed description can be found elsewhere.[29]

The simulations were performed on a model system that is embedded in a simple cubic lattice of volume $V = L \times L \times L$ with $L = 60$ and 72. Periodic boundary conditions were applied to all three directions. The system was composed of three components: AB diblock copolymers, a common solvent (S1), and a selective solvent (S2). The diblock copolymer chains used in the simulation were of the type $A_nB_{N-n}$, where $N$ is the total number of monomers and $n$ is the number of A monomers. In our simulations, the volume fraction of A monomer was fixed as $f_A = n/N = 3/4$, and two types of chains with chain lengths of $N = 8$ and $12$ were studied. The number of diblock copolymer chains in a system is denoted as $N_C$. Thus, the copolymer concentration is specified by $C_p = N_C N/V$. The initial configuration was generated by randomly creating $N_C$ chains of the model diblock copolymer, where each monomer occupied one lattice site and two consecutive monomers were connected by bonds that can adopt lengths of 1 or $\sqrt{2}$. Thus, each lattice site had 18 nearest neighbor sites. The copolymers were assumed to be self- and mutual-avoiding, meaning that no two monomers could occupy the same site simultaneously. After the desired number of copolymer chains had been generated, the unoccupied sites were designated as solvent S1 molecules, where each solvent molecule occupies one lattice site. Then, the character of randomly selected S1 molecules was changed to solvent S2 until the desired fraction $C_{S2}$ was reached.

The energy of the system was the objective function in the simulated annealing. In this simulation, only the nearest-neighbor interactions were considered, which were modeled by assigning an energy $E_{ij} = \varepsilon_{ij} k_B T_{ref}$ to each nearest-neighbor pair of unlike components $i$ and $j$, where $i, j$ = A, B, S1 (common solvent), and S2 (selective solvent) ($i \neq j$); $\varepsilon_{ij}$ is the reduced interaction energy; $k_B$ is the Boltzmann constant, and $T_{ref}$ is a reference temperature. It was assumed that $\varepsilon_{ii} = 0$, with $i$ = A, B and S1, S2. Immiscibility of the two blocks was ensured by setting $\varepsilon_{AB} = 1$. Solvent S1 was chosen as a common solvent, so it was assumed that $\varepsilon_{AS1} = \varepsilon_{BS1} = 0$. Solvent S2 was chosen as a selective solvent, so we set $\varepsilon_{AS2} > 0$ and $\varepsilon_{BS2} \leq 0$. These assignments ensured that solvent S2 was good for the B blocks and poor for the A blocks and that the A and B blocks were immiscible. In addition, the copolymer concentration $C_p$, S2 fraction $C_{S2}$, and the interaction parameters $\varepsilon_{AS2}$ and $\varepsilon_{BS2}$ were varied systemically to examine their effects on the self-assembled aggregates.

Each simulation started from a randomly generated initial configuration. Starting from the initial state, the ground state of the system was obtained by executing a set of Monte Carlo simulation at decreasing temperatures. Two types of trial moves were used in the simulations: chain overturning and exchange movement.[29] In a chain overturning move, a chain was selected, and all the A monomers on the chain exchanged sites with the corresponding B monomers on the same chain. In an exchange movement, there were three exchange types. (I) First, a monomer was selected randomly to exchange with a solvent molecule on one of its 18 nearest-neighbors. If the exchange did not break the chain, the exchange was allowed. If the exchange created a single break in the chain, the solvent molecule continued to exchange with subsequent monomer(s) along the broken chain until reconnection occurred. If the exchange broke the chain into more than two parts, the movement was not allowed. (II) A solvent molecule was selected randomly to exchange with a solvent molecule or a monomer on one of its 18 nearest-neighbors. If the selected neighbor was a monomer on a chain, then the movement was equivalent to type I. If the selected neighbor was a different type of solvent molecule, the exchange movement was allowed. (III) Two solvent molecules were selected randomly. If one of them is an S1 molecule and the other an S2 molecule, then the exchange movement was allowed. The acceptance or rejection of the attempted moves was further governed by the Metropolis rule.[39]

The annealing procedure followed a commonly used linear schedule as $T_j = fT_{j-1}$, where $T_j$ is the temperature used in the $j$th annealing step and $f$ is a scaling factor. Starting at an initial temperature, the annealing continued until the number of annealing steps reached a predetermined value. Specifically, the scaling factor $f$ was taken as 0.98 or 0.955, depending on the difference between the average energies of the system at the previous two annealing steps; $f = 0.955$ was used when the average energy difference was small, and $f = 0.98$ was used when the average energy difference was large. The initial temperature was $T_1 = 140T_{ref}$, and 140 annealing steps were performed. At each annealing step, 7500 Monte Carlo steps (MCS) were carried out. One MCS is defined as the average time taken for all the lattice sites to be visited for an attempted move. Simulations with several different random number generator seeds were performed to test the robustness of the observed self-assembled morphologies. Good reproducibility of the morphologies was obtained.

## III. RESULTS AND DISCUSSION

In this section, we present our simulation results for AB diblock copolymers in solutions of binary mixtures of solvents. The morphology diagrams are displayed in terms of the copolymer concentration and the selective solvent fraction (Figure 1). The average contact numbers and the mean square end-to-end distances of the chains are computed (Figures 2).The effect of the selectivity of the solvent on the morphology was investigated (Figures 3 and 4). The mechanism of morphological transitions is discussed by analyzing the variations of the average contact numbers between the A or the B monomers and the two types of solvent molecules, as well as the mean square end-to-end distances of the chains (Figures 2 and 5). The effect of chain length on the morphologies was also investigated (Figure 6). The vesicle sizes were computed based on the radial density profile of A monomers (Figure 7). The vesicle size variation is presented as a function of the selective solvent S2 fraction, the copolymer concentration, and the copolymer–solvent interaction (Figures 8–10), respectively.

A.  Morphology diagrams

We systematically varied the copolymer concentration $C_p$ and the fraction of selective solvent $C_{S2}$ to study their effect on the self-assembled microstructures for the model diblock copolymer $A_6B_2$, and the results are summarized in Figure 1, where the interaction parameters between the polymers and selective solvent S2 are set as $\varepsilon_{AS2} = 3$ and $\varepsilon_{BS2} = 0$. From Figure 1, it is noted that for $C_{S2} < 20\%$, the copolymers are dispersed in solution, where no aggregates are formed. This result indicates that the common solvent is dominant in the solution. The copolymers begin to form sphere-like aggregates at $C_{S2} = 20\%$ for systems with $C_p \geq 1.0\%$, and this copolymer concentration can be considered as the critical micelle content (CMC) of the selective solvent. Further, the morphologies of the aggregates depend on $C_p$. At $C_p = 1.0\%$, a wide range of rod-like aggregates were formed on increasing $C_{S2}$ to 30–35%, and at $C_{S2} = 40\%$, vesicles were formed. When $C_p = 1.5\%$, morphological sequences of a mixture of sphere-like and short-rod-like aggregates, rod-like aggregates, ring-shaped aggregates, sheet-like aggregates, and vesicles were formed at $C_{S2} = 25\%$, 30%, 32.5%, 35%, and 40%, respectively. When $C_p = 2.0\%$, the copolymers showed similar morphological sequences to those at $C_p = 1.5\%$, except that a bowl-like structure, which can be regarded as an unsealed vesicle, was formed instead of the sheet-like structure, and vesicle structures were also observed occasionally at $C_{S2} = 35\%$ as degenerated aggregates. When $C_p = 2.5\%$, ring-shaped aggregates were more frequently observed than rod-like aggregates. Besides the ring shaped aggregates, a new structure corresponding to cage-like aggregates was also found at $C_{S2} = 32.5\%$, just before vesicles formed. The ring-shaped aggregates can be regarded as two-dimensional structures, whereas the cage-like aggregates can be thought of as three-dimensional structures. Cage-like aggregates have also been reported by He and Schmid.[40] When $C_p \geq 3.0\%$, the morphological sequences were similar to those obtained for $C_p = 2.5\%$, except that the cage-like aggregates and the vesicles were formed at lower $C_{S2}$ values with increasing $C_p$. The mixture of sphere- and rod-like aggregates was replaced by a mixture of short-rod-like and rod-like aggregates, and vesicles were observed at $C_{S2} = 32.5\%$ when $C_p = 3.5\%$. When the copolymer concentration was higher ($C_p$ = 4.0% or 5.0%), vesicles formed when $C_{S2} \geq 32.5\%$. The cage-like structures were frequently obtained just before the vesicles formed as degenerated aggregates with a ring-like morphology. In dilute solution ($C_p$ = 0.5%), short rod-like aggregates were the dominant aggregates for $C_{S2}$ between 25% and 40%, except that sphere-like aggregates were observed to coexist with the short-rod aggregates at $C_{S2}$ = 25%.

It is well known that the self-assembly morphology of block copolymers is the result of competition between the interfacial energy and the entropy. To elucidate the mechanism of the morphological transformation, the average contact numbers for the monomers and the mean square end-to-end distance $Dee^2$ were computed, as plotted in Figure 2 as a function of $C_{S2}$ at $C_P$ = 1.5%. The average contact numbers for the A monomers with A monomers, B monomers, common solvent S1, and selective solvent S2 are defined as $N_{AA}$, $N_{AB}$, $N_{AS1}$, and $N_{AS2}$, respectively. Similarly, the average contact numbers for the B monomers with A monomers, B monomers, solvent S1, and solvent S2 are defined as $N_{BA}$, $N_{BB}$, $N_{BS1}$, and $N_{BS2}$, respectively. From Figure 2a, it is noted that with increasing $C_{S2}$, $N_{AB}$ and $N_{AS2}$ are almost unchanged, whereas $N_{AA}$ increases and $N_{AS1}$ decreases. From Figure 2b, it is noted that with increasing $C_{S2}$, $N_{BA}$ is almost unchanged, whereas $N_{BB}$ and $N_{BS2}$ increase, but $N_{BS1}$ decreases. Clearly, the decreases of $N_{AS1}$ and $N_{BS1}$ are due to the decreased amount of solvent S1 in the system, whereas the increase of $N_{BS2}$ is due to the increased amount of selective solvent S2. The increases of $N_{AA}$ and $N_{BB}$ indicate that both A and B monomers gradually aggregate more closely as the amount of S2 increases, which leads the contact of A–S2 close to zero. In the cases shown in Figures 2a and 2b, on increasing the amount of S2, the A–S2 and A–B contact is almost unchanged, whereas the A–S1 and B–S1 contact decreases and the B–S2 contact increases slightly. That is, increasing the amount of S2 decreases the total contact area between the micellar core (A-domain) and the solvents because the mixed solvent becomes poorer for the A-blocks, whereas the total contact area between the micellar corona (B-domain) and the solvents increases because the mixed solvent becomes better for the B-blocks. It has been deduced that at a constant volume, the total surface area always decreases when aggregates change from spheres to rods to vesicles,[29] which means that, with increasing $C_{S2}$, the interfacial energy part of the free energy decreases. From Figure 2c, it is noted that $Dee^2$ increases significantly when $C_{S2}$ increases from 15% to 35% as the copolymers from the dissolved state transform to spheres, rods, rings, and sheet-like aggregates. For vesicle morphologies, the variation in $Dee^2$ is relatively small. The increase of $Dee^2$ means that the copolymer chains become increasingly stretched. The number of available conformations decreases with increasing $Dee^2$, which leads to a decrease in the chain conformational entropy and an increase in the conformational part of the free energy. Therefore, the basic morphologies of spheres to rods to vesicles obtained by increasing the amount of S2 (Figure 1) are the result of competition between the interfacial energy and the chain conformational entropy. In contrast, the sheet-like, ring-like, and cage-like aggregate morphologies are all intermediate between the rods and the vesicles. On the other hand, it is noted that the $N_{AS1}$ value is always larger than 6.0, indicating that a large amount of solvent molecules are in contact with the solvophobic blocks.

The selectivity of the selective solvent may play an important role in microstructure formation of amphiphilic macromolecules. To investigate this effect, we fixed the concentration of the diblock copolymer $A_6B_2$ at $C_p = 2.5\%$ and studied the morphology variation on increasing both $\varepsilon_{AS2}$ and the fraction of the selective solvent $C_{S2}$ for two cases ($\varepsilon_{BS2} = 0$ or -1).

Figure 3 exhibits the morphological diagram as a function of $\varepsilon_{AS2}$ and $C_{S2}$ when $\varepsilon_{BS2} = 0$.

Compared with the case presented in Figure 1 where $\varepsilon_{AS2} = 3$, it is noted that the $\varepsilon_{AS2}$ value affects the structure formation. When $\varepsilon_{AS2} = 1$, rod-like and sheet-like aggregates are the dominate morphologies, whereas vesicles do not appear, even at high $C_{S2}$ values. On increasing $\varepsilon_{AS2}$ to 2–3, rod-like aggregates transform into bowl-like vesicles at $C_{S2} = 35\%$ and vesicles are formed at $C_{S2} = 40\%$. When $\varepsilon_{AS2} = 4$–5, vesicles appear at $C_{S2} = 35\%$. It is also noted that vesicles do not appear when $C_{S2} < 35\%$, even when the repulsive strength is high with $\varepsilon_{AS2} > 3$.

Figure 4 presents the morphological diagram as a function of $\varepsilon_{AS2}$ and $C_{S2}$ when $\varepsilon_{BS2} = -1$. A comparison of the diagrams shown in Figures 3 and 4 reveals significant morphology changes. Unlike the rich morphologies shown in Figure 3, the stacked lamellae aggregates are dominant in the diagram shown in Figure 4, and vesicles only appear in a small region with both larger $\varepsilon_{AS2}$ and $C_{S2}$ values. From the diagram in Figure 4, it is noted that the composition of the outside layers of a stacked lamellar aggregate changes with $C_{S2}$. When $C_{S2}$ is smaller, the outside layers are composed of the A-blocks only, whereas when $C_{S2}$ is larger, the outside layers are composed of the B-block only. Between these two cases, the outside layers contain both A- and B-blocks, and they can be located on either side of one aggregate or in different aggregates in the system. Based on the differences in the composition of the outside layers and the morphology, the diagram in Figure 4 can be divided into different parts using straight lines. Notably, with increasing $\varepsilon_{AS2}$, the line dividing the different parts shifts to lower $C_{S2}$ values.

To elucidate the mechanism for formation of the stacked lamellae, a sequence of morphologies of $A_6B_2$ were examined by increasing the selective strength for the B-block $\varepsilon_{BS2}$ from 0 to -1, as shown in Figure 5a, where $\varepsilon_{AS2} = 1$ and $C_{S2} = 30\%$. In the weak $\varepsilon_{BS2}$ range, only spheres and rods are observed. When $\varepsilon_{BS2} = -0.1$, disk-like aggregate appear, which coexist with the spheres. For stronger selective strengths, only disk-like aggregates are formed. The number of stacked lamellae in each micelle increases with increasing the selective strength. Moreover, a large worm-like aggregate of stacked lamellae is formed in the stronger selective strength range of $\varepsilon_{BS2}$ = -0.5 to -1. The average contact numbers for the monomers with monomers and solvent molecules were computed, as plotted in Figures 5b and 5c as a function of $\varepsilon_{BS2}$. From Figure 5b, it is noted that on changing $\varepsilon_{BS2}$ from 0 to -1.0, $N_{AB}$ and $N_{AS2}$ are almost unchanged, whereas $N_{AA}$ increases and $N_{AS1}$ decreases. From Figure 5c, it is noted that on changing $\varepsilon_{BS2}$ from 0 to -1.0, $N_{BA}$ is almost unchanged, but $N_{BS2}$ increases and $N_{BS1}$ and $N_{BB}$ decrease. In particular, the increase of $N_{BS2}$ and the decrease of $N_{BS1}$ are rapid in the range of $\varepsilon_{BS2}$ = 0 to -0.3. The larger increase of $N_{BS2}$ is due to increased attractive interactions between the B monomers and S2, which cause a corresponding decrease of $N_{BB}$. When the morphology changes from the rod-like aggregate to a stacked lamellar aggregate, contact between the solvophobic block and the solvents decreases slightly, whereas the contact between the solvophilic block and the solvents increases considerably. That is, the stacked lamellar aggregate allows a large contact between the solvophilic block and the solvents, which deceases the interfacial energy of the system. From Figure 5d, it is noted that $Dee^2$ increases significantly when $\varepsilon_{BS2}$ changes from $\varepsilon_{BS2}$ = 0 to -0.3, and the copolymers in spheres transform to rods and further to stacked lamellar aggregates. The $Dee^2$ curve in Figure 5d and the $N_{BS2}$ curve in Figure 5c have similar trends. As stated earlier, the increase of $Dee^2$ leads to a decrease in the chain conformational entropy and an increase in the conformational part of the free energy. Therefore, the morphological transition from spheres to rods to stacked lamellar aggregates observed in Figure 5a on changing $\varepsilon_{BS2}$ are also the result of competition between the interfacial energy and the chain conformational entropy.

The above simulation results can be compared with related experiments or simulations. The morphological transition sequence (Figure 1) with increasing $C_{S2}$ at a fixed $C_P$ value, from dissolved chains, to sphere-like aggregates, to short-rod-like aggregates, to rod-like aggregates, to ring-/cage-like aggregates, and finally to vesicles is consistent with that observed experimentally by Eisenberg's group.[21,22] This morphological sequence with increasing $C_{S2}$ at a fixed $C_P$ value is also consistent with that predicted by simulation based on only one type of solvent but change the property of the solvent.[29] The stacked lamellar aggregates predicted in our simulations are consistent with those observed experimentally by Eisenberg's group as small lamellea.[21] Using self-consistent field theory, Xia et al. predicted anisotropic ellipsoidal micelles composed of segmented layers of A/B domains in their study of symmetric AB diblock copolymers in C-homopolymers (or solvents).[41] These anisotropic ellipsoidal micelles are similar to the stacked lamellar aggregates predicted in our study. On the other hand, our finding that there are large amounts of solvent molecules in contact with the solvophobic blocks in all the aggregates is completely different from that predicted by simulations based on only one type of solvent.[29]

B. Effect of chain length on morphology

The effect of chain length on morphology transitions was also investigated, with a focus on comparing the morphologies formed by diblocks $A_9B_3$ and $A_6B_2$, which have the same volume fraction $f_A$ and the same interaction parameters. Because of the longer chain length, larger simulation boxes with $L = 72$ were used for $A_9B_3$ systems. We chose a polymer concentration of $C_p = 2.5\%$ and varied $C_{S2}$ from 10% to 50%. A morphological sequence from dissolved chains, to spherical-like aggregates, to short rod-like aggregates, to rod-like aggregates, to ring-like aggregates, and finally to vesicles was obtained with increasing $C_{S2}$, as shown in Figure 6. This sequence is similar to that determined for $A_6B_2$ in the same parameter space, as shown in Figure 1 at $C_p = 2.5\%$. When compared with those for $A_6B_2$, the positions of the morphology transitions for $A_9B_3$ were shifted to lower $C_{S2}$ values, indicating that to form the same structure, the $A_9B_3$ copolymer needs less selective solvent than the $A_6B_2$ copolymer at the same copolymer concentration.

C. Analysis of vesicle structure

To analyze the vesicle structure, the radial density profiles of the A and B monomers and the solvents were calculated. Variations in the average density profiles of monomers A and B and solvents S1 and S2 with $r$ for a typical vesicle structure formed by $A_9B_3$ at $C_{S2} = 35\%$ are plotted in Figure 7, where $r$ is the distance from the center of mass of the vesicle. It is noted that in the central part of the vesicle, there is only solvent, including both types of solvent molecules. Moreover, the common solvent is also found inside the vesicle shell formed by the A monomers. The density peak of the B monomer at the outer surface is obviously lower than that at the inner surface, which is consistent with the results obtained by Du et al.[42]

From the density profile of the monomers, we obtained the shell thickness of the vesicle by calculating the half-height width in the density curve of the A monomer. The schematic plot in Figure 8a shows the definitions of the outer and inner radius $R_{out}$ and $R_{in}$ and the thickness $\Delta R$ of a vesicle. Figure 8b shows the variations of $R_{out}$, $R_{in}$, and $\Delta R$ with $C_{S2}$ for $A_9B_3$ in the range of $C_{S2} = 32.5–50\%$. It is noted that with increasing $C_{S2}$, $\Delta R$ increases slowly; in contrast, both $R_{out}$ and $R_{in}$ decrease. With increasing $C_{S2}$, the effective repulsive interaction of the solvents with the micelle core increases, hence $R_{out}$ decreases and some solvent molecules inside the vesicle are pushed out. As a result, the vesicle shrinks, and it can be deduced that $R_{in}$ shrinks more than $R_{out}$ does. Therefore, the shell thickness $\Delta R$ increases with increasing $C_{S2}$.

We also investigated the effect of copolymer concentration $C_p$ on the vesicle size, and the results for the $A_6B_2$ system with increasing $C_P$ from 1% to 5% at $C_{S2} = 40\%$ are plotted in Figure 9. The curve in Figure 9 can be divided into two parts in different ranges of $C_p$: (I) 1.0–2.5% and (II) 3.0–5.0%, where the curves in each part increases almost linearly with increasing $C_p$. For both $R_{out}$ and $R_{in}$, the slope in part II is slightly smaller than the corresponding value in part I. It can be easily deduced that when the same amount of copolymer is added to a vesicle, the increase in volume is larger for a smaller vesicle than for a bigger vesicle. This is why the slope in part II is slightly smaller than the corresponding value in part I. On the other hand, in part I the slope of $R_{in}$ is slightly smaller than that of $R_{out}$, whereas in part II the slope of $R_{in}$ is much smaller than that of $R_{out}$. Therefore, the shell thickness $\Delta R$ of vesicles increases slowly in part I and much more quickly in part II. The tendency for the vesicle size to increase with increasing copolymer concentration was also observed in Eisenberg's study.[22] However, there is one difference between our results and theirs. In their case, the variation of the vesicle size is steeper when the copolymer concentration is higher (3–5%) than that when $C_P$ is lower (0.6–3%), resulting in an almost constant shell thickness. In our case, the shell thickness increases with increasing $C_P$, and the slope of $\Delta R$ is smaller in part I than in part II.

We also investigated the effect of $\varepsilon_{AS2}$ on the vesicle size, and the results for $C_{S2} = 40\%$ are plotted in Figure 10. It is noted that the vesicle size does not change considerably with increasing $\varepsilon_{AS2}$. Compared with the variations observed in Figures 8b and 9, it is obvious that the effects of $C_{S2}$ and $C_P$ on vesicle size are more pronounced than that of $\varepsilon_{AS2}$.

## IV. CONCLUSION

We have systematically investigated the self-assembly of amphiphilic AB diblock copolymers in mixtures of a common solvent (S1) and a B-block selective solvent (S2) using a simulated annealing method applied to a lattice model of block copolymers. Phase diagrams for the copolymers were constructed by varying the copolymer concentration, the fraction of selective solvent, and the selectivity of the selective solvent. Rich phase transition sequences, e.g., disorder → sphere → rod →ring → vesicle, were observed, which are consistent with that observed experimentally by Eisenberg's group.[21,22] The selectivity of the selective solvent, characterized by both $\varepsilon_{AS2}$ and $\varepsilon_{BS2}$, affected the self-assembled microstructure significantly. In particular, when the interaction $\varepsilon_{BS2}$ was negative, stacked lamellae aggregates were the dominant species in the phase diagram. The outside layers of the stacked lamellae may be composed of A-blocks only, B-blocks only, or both A- and B-blocks. The mechanisms of aggregate transformation and the formation of stacked lamellar aggregates were discussed by analyzing the variations of the average contact numbers of the A or B monomers with monomers and with the molecules of the two solvent types, as well as the mean square end-to-end distance of chains. It was found that the basic morphological sequence of spheres to rods to vesicles led to a decrease in the interfacial energy, but an increase of $Dee^2$, and hence a decrease in the chain conformational entropy and an increase in the conformational part of the free energy. The stacked lamellar aggregates allowed greater contact between the solvophilic block and the solvents and had a larger $Dee^2$ value. Therefore, it is the competition between the interfacial energy and the chain conformational entropy that results in the basic morphological sequence and the formation of stacked lamellar aggregates. An investigation of the effect of chain length on the morphology transitions revealed that to form the same structure, a longer copolymer needs less selective solvent than that a shorter copolymer with the same copolymer concentration and volume fraction. Analysis of the vesicle structure showed that the vesicle size increased with increasing $C_P$ or with decreasing $C_{S2}$, but remained almost constant with changes in $\varepsilon_{AS2}$.

*Acknowledgements:* This work was supported by the National Natural Science Foundation of China (21204040, 20904027, 21574071, 21528401, 20925414, and 91227121), by PCSIRT (IRT1257), and by the 111 Project (B16027).

**References**

1. S. A. Jenekhe and X. L. Chen, Self-assembled aggregates of rod-coil block copolymers and their solubilization and encapsulation of fullerenes, *Science* 279(5358), 1903 (1998) http://dx.doi.org/10.1126/science.279.5358.1903

2. R. Stoenescu, A. Graff, and W. Meier, Asymmetric ABC-triblock copolymer membranes induce a directed insertion of membrane protein, *Macromol. Biosci.* 4(10), 930 (2004) http://dx.doi.org/10.1002/mabi.200400065

3. R. Stoenescu and W. Meier, Vesicles with asymmetric membranes from amphiphilic ABC triblock copolymers, *Chem. Commun.* 24(24), 3016 (2002) http://dx.doi.org/10.1039/b209352a

4. Y. F. Zhou and D. Y. Yan, Real-time membrane fusion of giant polymer vesicles, *J. Am. Chem. Soc.* 127(30), 10468 (2005) http://dx.doi.org/10.1021/ja0505696

5. Y. F. Zhou and D. Y. Yan, Real-time membrane fission of giant polymer vesicles, *Angew. Chem. Int. Ed.* 44(21), 3223 (2005) http://dx.doi.org/10.1002/anie.200462622

6. R. Savic, L. B. Luo, A. Eisenberg, and D. Maysinger, Micellar nanocontainers distribute to defined cytoplasmic organelles, *Science* 300(5619), 615 (2003) http://dx.doi.org/10.1126/science.1078192

7. C. Allen, D. Maysinger, and A. Eisenberg, Nano-engineering block copolymer aggregates for drug delivery, *Colloids Surf. B Biointerfaces* 16(1-4), 3 (1999) http://dx.doi.org/10.1016/S0927-7765(99)00058-2

8. L. F. Zhang and A. Eisenberg, Multiple morphologies of "crew-cut" aggregates of polystyrene-*b*-poly(acrylic acid) block copolymers, *Science* 268(5218), 1728 (1995) http://dx.doi.org/10.1126/science.268.5218.1728

9. X. T. Shuai, H. Ai, N. Nasongkla, S. Kim, and J. M. Gao, Micellar carriers based on block copolymers of poly(ε-caprolactone) and poly(ethylene glycol) for doxorubicin delivery, *J. Control. Release* 98(3), 415 (2004) http://dx.doi.org/10.1016/j.jconrel.2004.06.003

10. H. Lomas, I. Canton, S. MacNeil, J. Du, S. P. Armes, A. J. Ryan, A. L. Lewis, and G. Battaglia, Biomimetic pH sensitive polymersomes for efficient DNA encapsulation and delivery, *Adv. Mater.* 19(23), 4238 (2007) http://dx.doi.org/10.1002/adma.200700941

11. X. B. Xiong, H. Uludag, and A. Lavasanifar, Biodegradable amphiphilic poly(ethylene oxide)-block-polyesters with grafted polyamines as supramolecular nanocarriers for efficient siRNA delivery, *Biomaterials* 30(2), 242 (2009) http://dx.doi.org/10.1016/j.biomaterials.2008.09.025

12. A. Blanazs, S. P. Armes, and A. J. Ryan, Self-assembled block copolymer aggregates: From micelles to vesicles and their biological applications, *Macromol. Rapid Commun.* 30(4-5), 267 (2009) http://dx.doi.org/10.1002/marc.200800713

13. K. J. Hanley, T. P. Lodge, and C. I. Huang, Phase behavior of a block copolymer in solvents of varying selectivity, *Macromolecules* 33(16), 5918 (2000) http://dx.doi.org/10.1021/ma000318b

14. C. Lai, W. B. Russel, and R. A. Register, Phase behavior of Styrene–Isoprene diblock copolymers in strongly selective solvents, *Macromolecules* 35(3), 841 (2002) http://dx.doi.org/10.1021/ma011696z

15. T. P. Lodge, B. Pudil, and K. J. Hanley, The full phase behavior for block copolymers in solvents of varying selectivity, *Macromolecules* 35(12), 4707 (2002) http://dx.doi.org/10.1021/ma0200975

16. B. Yu, B. Li, P. Sun, T. Chen, Q. Jin, D. Ding, and A. C. Shi, Cylinder-gyroid-lamella transitions in diblock copolymer solutions: A simulated annealing study, *J. Chem. Phys.* 123(23), 234902 (2005) http://dx.doi.org/10.1063/1.2137711

17. T. Suo, D. Yan, S. Yang, and A. C. Shi, A theoretical study of phase behaviors for diblock copolymers in selective solvents, *Macromolecules* 42(17), 6791 (2009) http://dx.doi.org/10.1021/ma900939u

18. M. Antonietti and S. Forster, Vesicles and liposomes: A self-assembly principle beyond lipids, *Adv. Mater.* 15(16), 1323 (2003) http://dx.doi.org/10.1002/adma.200300010

19. D. J. Pochan, Z. Y. Chen, and H. G. Cui, Toroidal triblock copolymer assemblies, *Science* 306(5693), 94 (2004) http://dx.doi.org/10.1126/science.1102866

20. X. S. Wang, G. Guerin, H. Wang, Y. S. Wang, I. Manners, and M. A. Winnik, Cylindrical block copolymer micelles and co-micelles of controlled length and architecture, *Science* 317(5838), 644 (2007) http://dx.doi.org/10.1126/science.1141382

21. Y. Y. Mai and A. Eisenberg, Self-assembly of block copolymers, *Chem. Soc. Rev.* 41(18), 5969 (2012) http://dx.doi.org/10.1039/c2cs35115c

22. H. W. Shen and A. Eisenberg, Morphological phase diagram for a ternary system of block copolymer $PS_{310}$-b-$PAA_{52}$/Dioxane/$H_2O$, *J. Phys. Chem. B* 103(44), 9473 (1999) http://dx.doi.org/10.1021/jp991365c

23. B. Du, A. Mei, K. Yin, Q. Zhang, J. Xu, and Z. Fan, Vesicle formation of $PLA_x$–$PEG_{44}$ diblock copolymers, *Macromolecules* 42(21), 8477 (2009) http://dx.doi.org/10.1021/ma9016339

24. H. Huang, B. Chung, J. Jung, H. W. Park, and T. Chang, Toroidal micelles of uniform size from diblock copolymers, *Angew. Chem. Int. Ed.* 48(25), 4594 (2009) http://dx.doi.org/10.1002/anie.200900533

25. A. G. Denkova, P. H. H. Bomans, M.O. Coppens, N. A. J. M. Sommerdijk, and E. Mendes, Complex morphologies of self-assembled block copolymer micelles in binary solvent mixtures: The role of solvent–solvent correlations, *Soft Matter* 7(14), 6622 (2011) http://dx.doi.org/10.1039/c1sm05461a

26. S. Zhong, H. Cui, Z. Chen, K. L. Wooley, and D. J. Pochan, Helix self-assembly through the coiling of cylindrical micelles, *Soft Matter* 4(1), 90 (2008) http://dx.doi.org/10.1039/B715459C

27. D. H. Han, X. Y. Li, S. Hong, H. Jinnai, and G. J. Liu, Morphological transition of triblock copolymer cylindrical micelles responding to solvent change, *Soft Matter* 8(7), 2144 (2012) http://dx.doi.org/10.1039/c2sm07020k

28. B. E. McKenzie, J. F. de Visser, H. Friedrich, M. J. M. Wirix, P. H. H. Bomans, G. de With, S. J. Holder, and N. A. J. M. Sommerdijk, Bicontinuous nanospheres from simple amorphous amphiphilic diblock copolymers, *Macromolecules* 46(24), 9845 (2013) http://dx.doi.org/10.1021/ma4019729

29. P. Sun, Y. Yin, B. Li, T. Chen, Q. Jin, D. Ding, and A.C. Shi, Simulated annealing study of morphological transitions of diblock copolymers in solution, *J. Chem. Phys.* 122(20), 204905 (2005) http://dx.doi.org/10.1063/1.1924452

30. R. Wang, Z. Jiang, and G. Xue, Excluded volume effect on the self-assembly of amphiphilic AB diblock copolymer in dilute solution, *Polymer* 52(10), 2361 (2011) http://dx.doi.org/10.1016/j.polymer.2011.03.010

31. J. Cui and W. Jiang, Vesicle formation and microphase behavior of amphiphilic ABC triblock copolymers in selective solvents: A Monte Carlo study, *Langmuir* 26(16), 13672 (2010) http://dx.doi.org/10.1021/la102211d

32. J. Cui and W. Jiang, Structure of ABCA tetrablock copolymer vesicles and their formation in selective solvents: A Monte Carlo study, *Langmuir* 27(16), 10141 (2011) http://dx.doi.org/10.1021/la202377t

33. J. Cui, Y. Y. Han, and W. Jiang, Asymmetric vesicle constructed by AB/CB diblock copolymer mixture and its behavior: A Monte Carlo study, *Langmuir* 30(30), 9219 (2014) http://dx.doi.org/10.1021/la501674a

34. P. T. He, X. J. Li, M. G. Deng, T. Chen, and H. J. Liang, Complex micelles from the self-assembly of coil-rod-coil amphiphilic triblock copolymers in selective solvents, *Soft Matter* 6(7), 1539 (2010) http://dx.doi.org/10.1039/b926370e

35. W. Kong, B. Li, Q. Jin, D. Ding, and A. C. Shi, Helical vesicles, segmented semivesicles, and noncircular bilayer sheets from solution-state self-assembly of ABC Miktoarm star terpolymers, *J. Am. Chem. Soc.* 131(24), 8503 (2009) http://dx.doi.org/10.1021/ja900405r

36. C. I. Huang and Y. C. Hsu, Effects of solvent immiscibility on the phase behavior and microstructural length scales of a diblock copolymer in the presence of two solvents, *Phys. Rev. E* 74(5), 051802 (2006) http://dx.doi.org/10.1103/PhysRevE.74.051802

37. I. Carmesin and K. Kremer, The bond fluctuation method: A new effective algorithm for the dynamics of polymers in all spatial dimensions, *Macromolecules* 21(9), 2819 (1988) http://dx.doi.org/10.1021/ma00187a030

38. R. G. Larson, Self-assembly of surfactant liquid crystalline phases by Monte Carlo simulation, *J. Chem. Phys.* 91(4), 2479 (1989) http://dx.doi.org/10.1063/1.457007 , R. G. Larson, Monte Carlo simulation of microstructural transitions in surfactant systems, *J. Chem. Phys.* 96(11), 7904 (1992) http://dx.doi.org/10.1063/1.462343

39. N. Metropolis, A. W. Rosenbluth, M. N. Rosenbluth, A. H. Teller, and E. Teller, Equation of state calculations by fast computing machines, *J. Chem. Phys.* 21(6), 1087 (1953) http://dx.doi.org/10.1063/1.1699114

40. X. He and F. Schmid, Spontaneous formation of complex micelles from a homogeneous solution, *Phys. Rev. Lett.* 100(13), 137802 (2008) http://dx.doi.org/10.1103/PhysRevLett.100.137802

41. B. Xia, W. Li, and F. Qiu, Self-assembling behaviors of symmetric diblock copolymers in C homopolymers, *Acta Chimi. Sin.* 72(1), 30 (2014) http://dx.doi.org/10.6023/A13080892

42. H. Du, J. Zhu, and W. Jiang, Study of controllable aggregation morphology of ABA amphiphilic triblock copolymer in dilute solution by changing the solvent property, *J. Phys. Chem. B* 111(8), 1938 (2007) http://dx.doi.org/10.1021/jp067221x

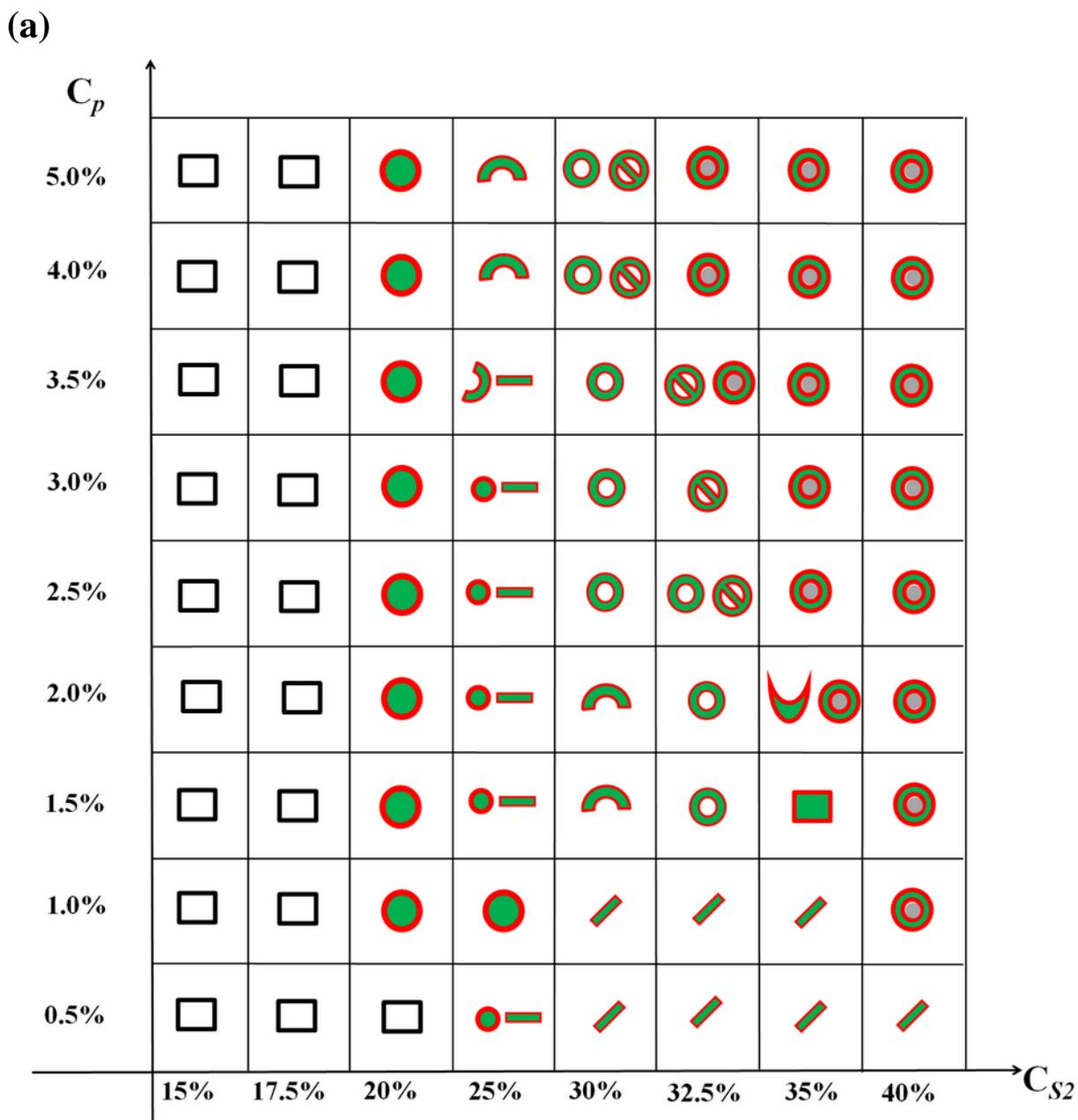

**Figure 1** Morphological diagram of diblock copolymer $A_6B_2$ in terms of concentration parameter $C_{S2}$ and copolymer concentration $C_p$ ($\varepsilon_{AS1} = \varepsilon_{BS1} = 0$, $\varepsilon_{AS2} = 3$, and $\varepsilon_{BS2} = 0$). (a) Morphological diagram and (b) morphology scheme (green: A monomer; red: B monomer; yellow: common solvent S1; grey: selective solvent S2. The cage structure given the density morphology of A monomer. The bowl and vesicle profiles correspond to cross sections of cut structures.).

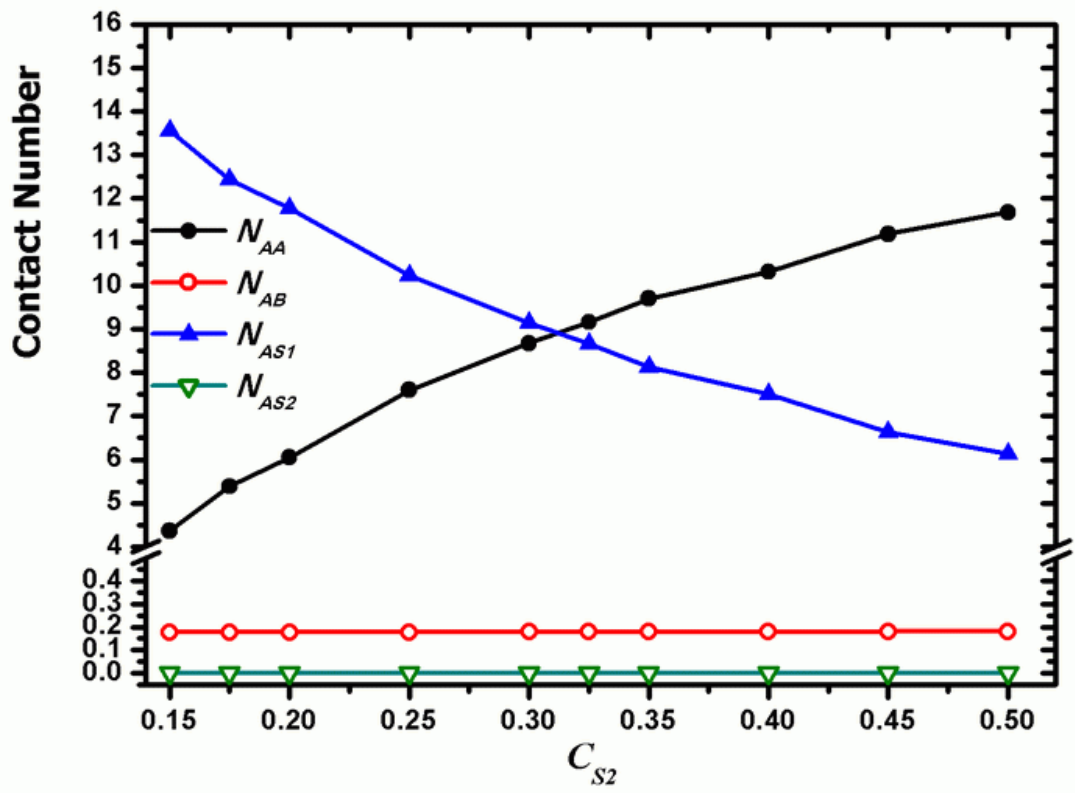

(a)

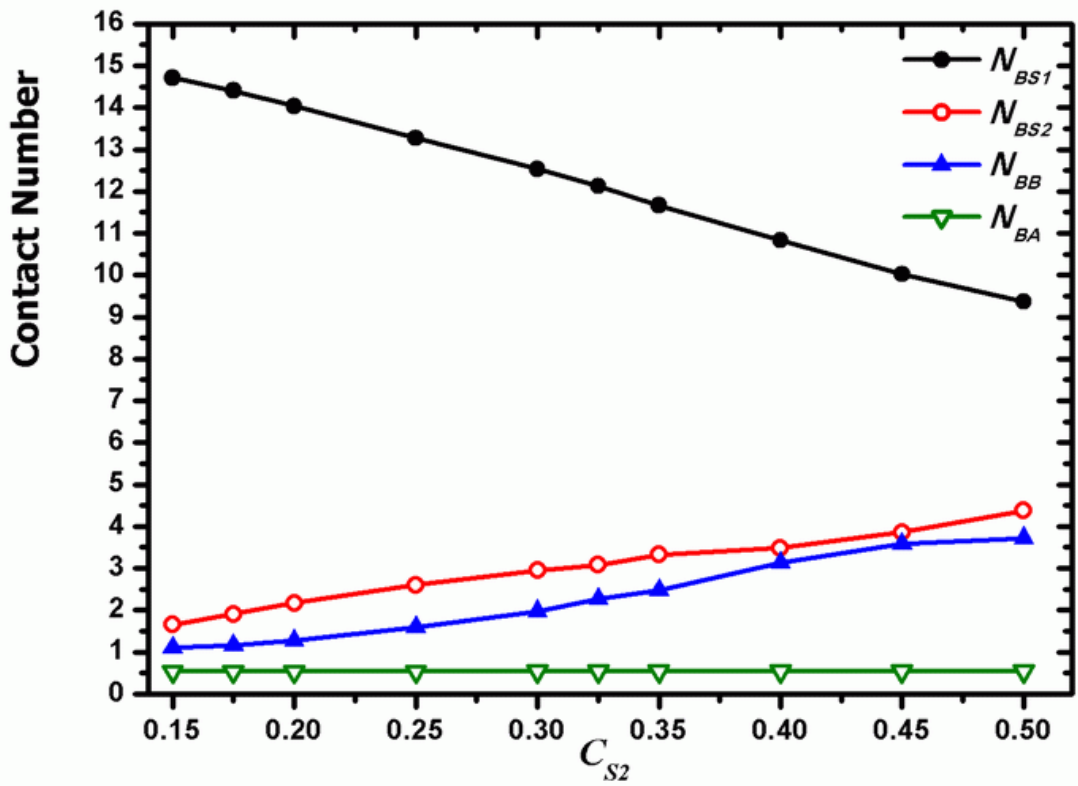

(b)

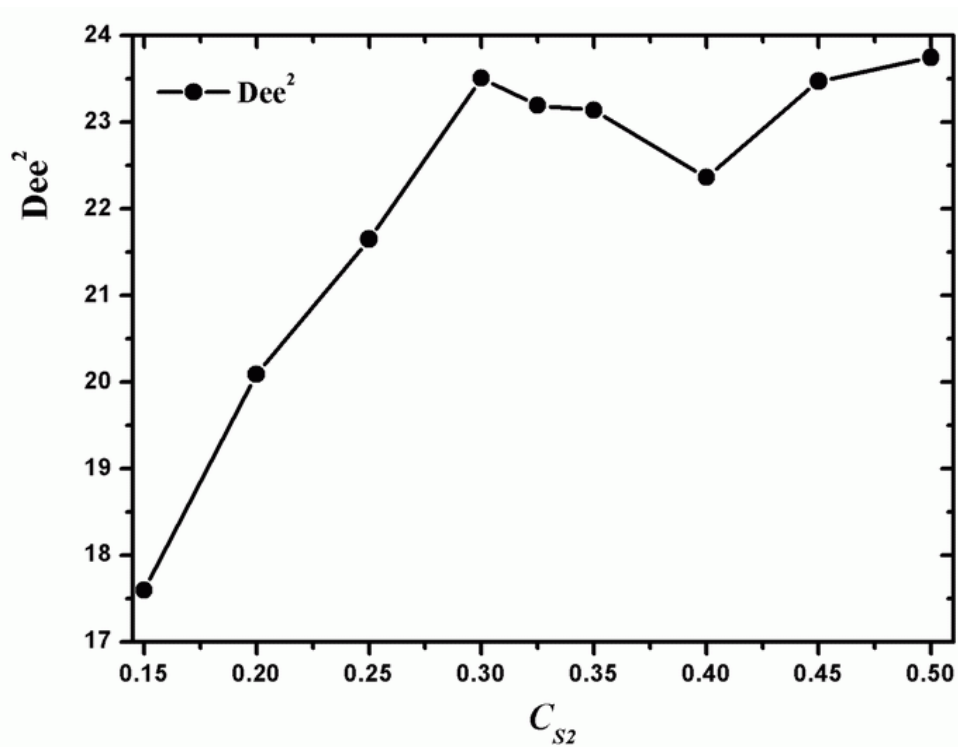

**Figure 2** Contact number and mean square end-to-end distance of diblock $A_6B_2$ as a function of selective solvent fraction $C_{S2}$ at $C_p = 1.5\%$ ($\varepsilon_{AS1} = \varepsilon_{BS1} = 0$, $\varepsilon_{AS2} = 3$, and $\varepsilon_{BS2} = 0$). (a) Contact numbers for the A monomer; (b) contact numbers for the B monomer; and (c) mean square end-to-end distance.

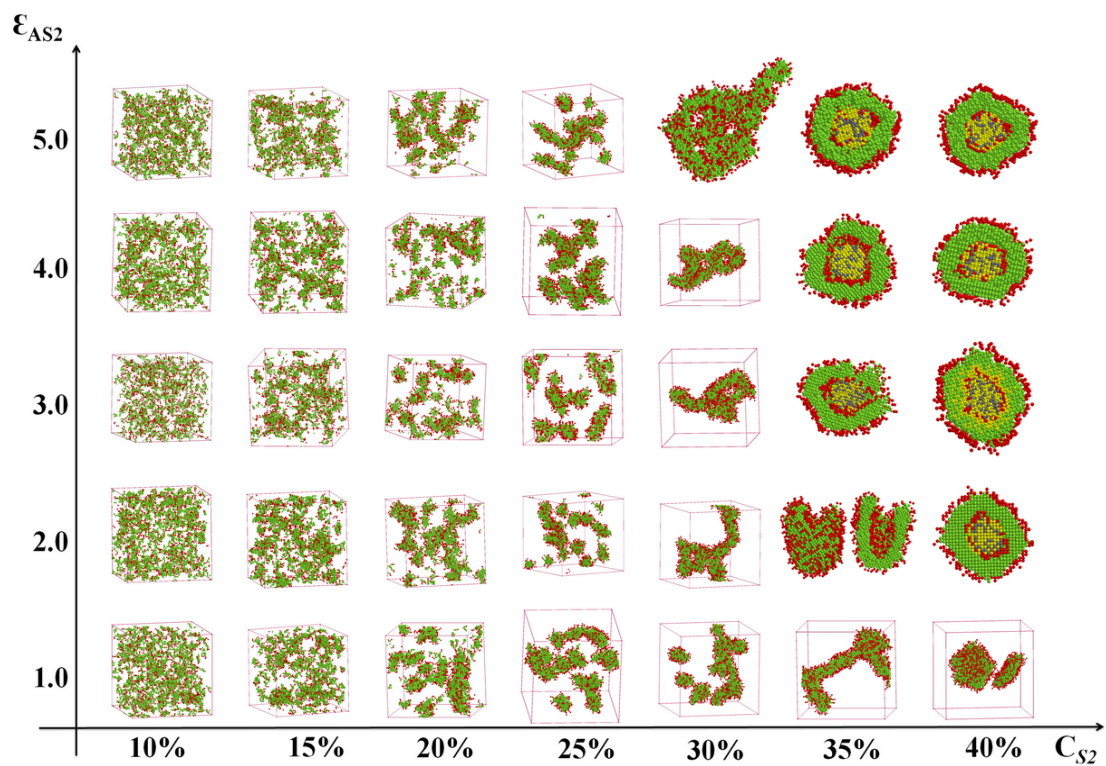

**Figure 3** Morphological diagram of $A_6B_2$ in terms of repulsive strength parameter $\varepsilon_{AS2}$ and selective solvent S2 fraction $C_{S2}$ at $C_P = 2.5\%$ for $\varepsilon_{BS2} = 0$ (green: A monomer; red: B monomer; yellow: common solvent S1; grey: selective solvent S2. The bowl and vesicle profiles correspond to cross sections of cut structures.)

(a)

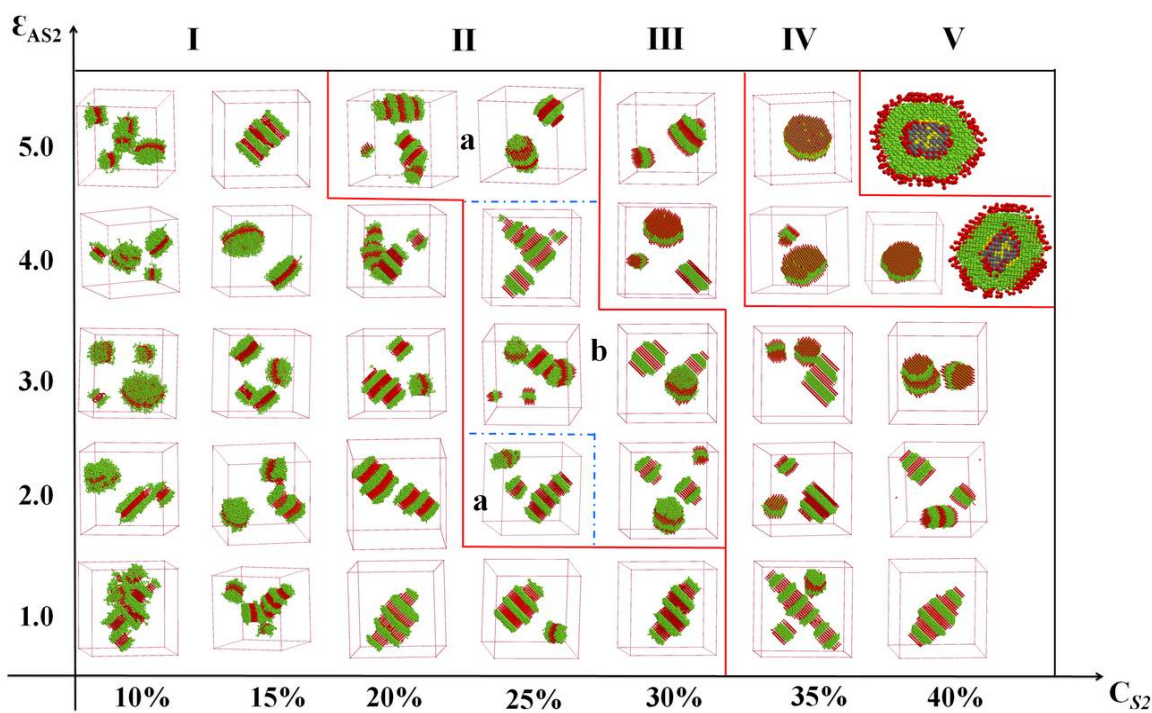

**Figure 4** Morphological diagram of $A_6B_2$ in terms of repulsive strength parameter $\varepsilon_{AS2}$ and selective solvent S2 fraction $C_{S2}$ at $C_P = 2.5\%$ for $\varepsilon_{BS2} = -1$. (Green: A monomer; red: B monomer; yellow: common solvent S1; grey: selective solvent S2. The bowl and vesicle profiles correspond to cross sections of cut structures.)

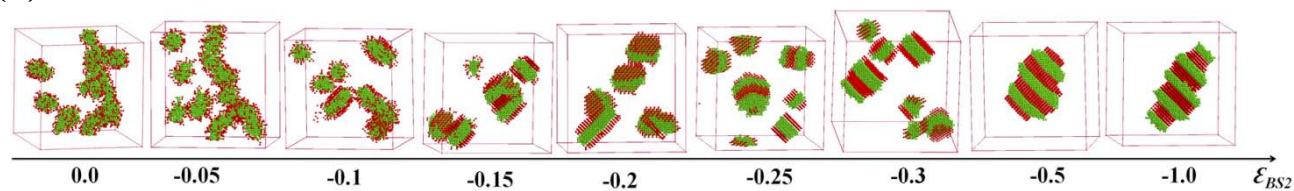

(b)

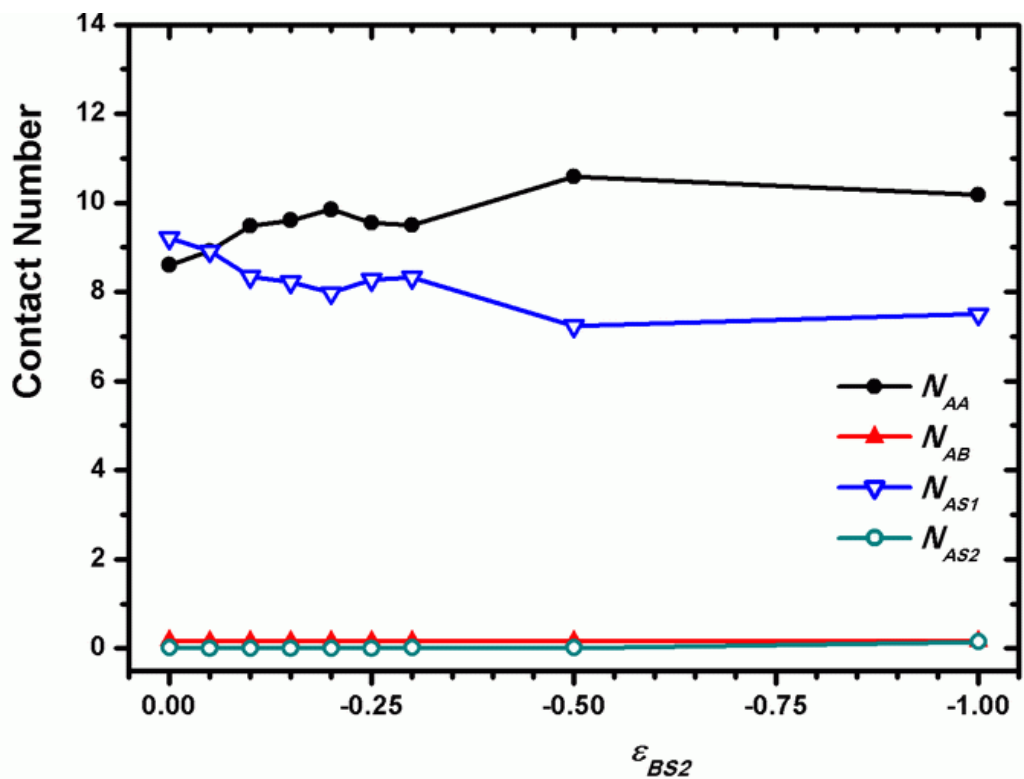

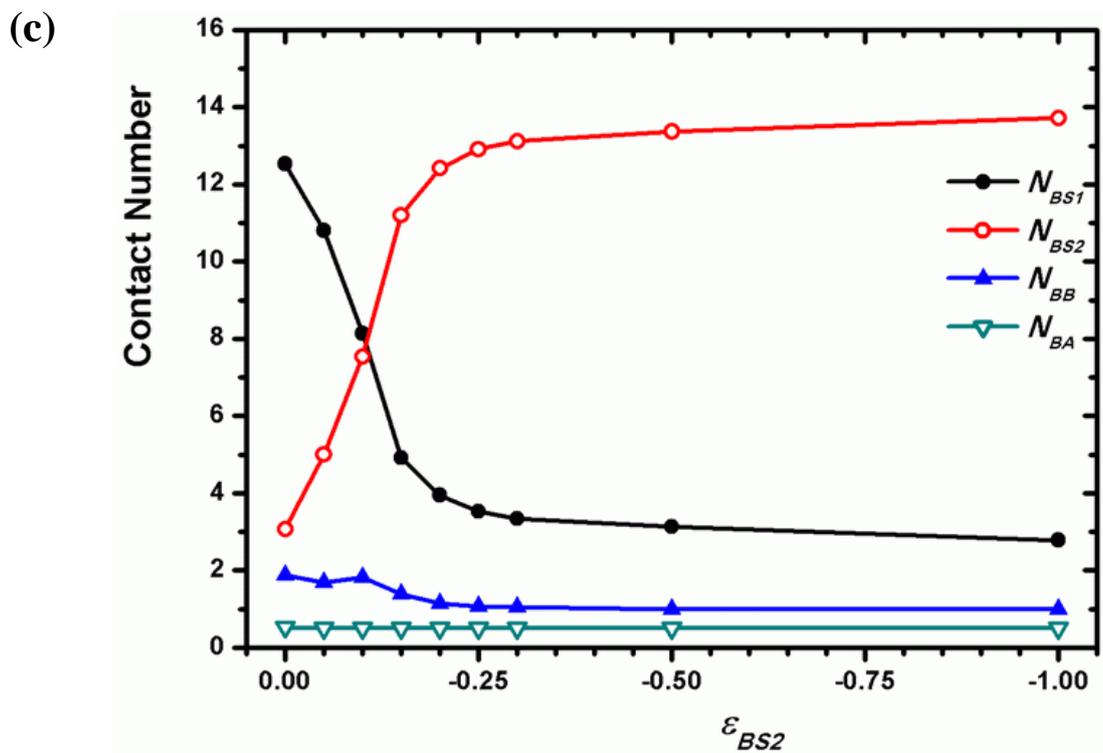

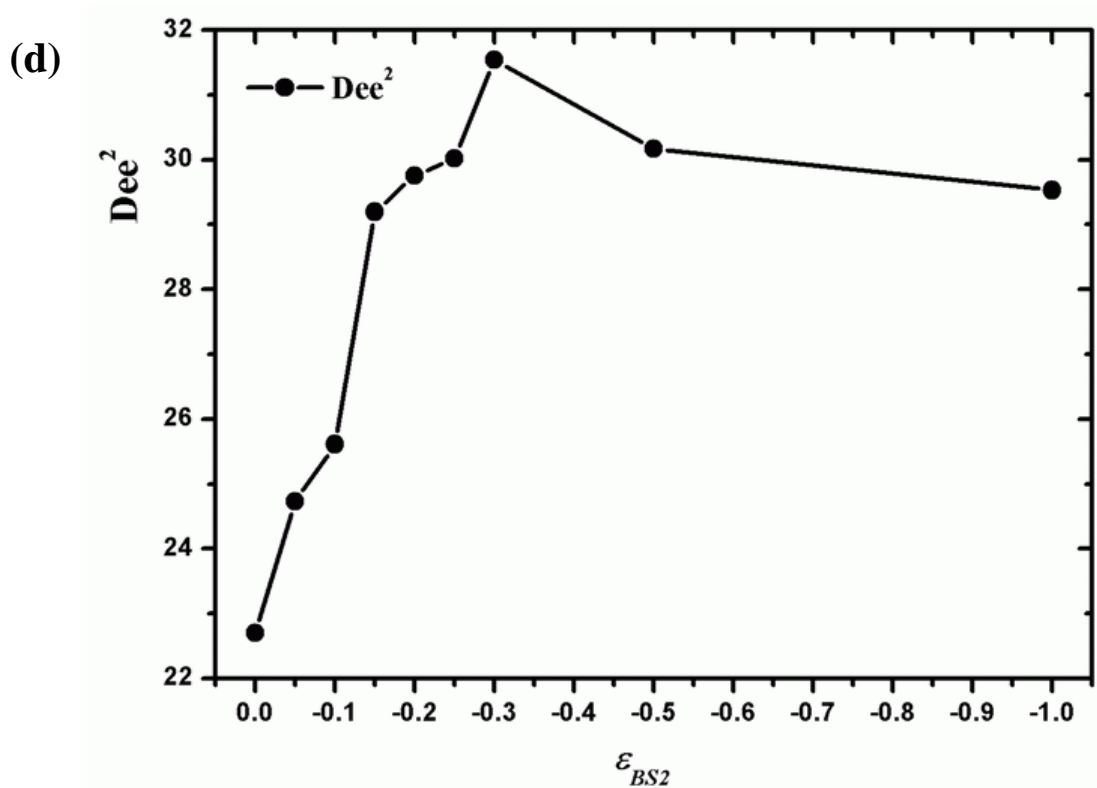

**Figure 5** (a) Morphology sequence of diblock $A_6B_2$ as a function of $\varepsilon_{BS2}$ at $C_p = 2.5\%$, $C_{S2} = 30\%$ ($\varepsilon_{AS1} = \varepsilon_{BS1} = 0$, $\varepsilon_{AS2} = 1$); (b) contact numbers for the A monomer; (c) contact numbers for the B monomer; and (d) the mean square end-to-end distance.

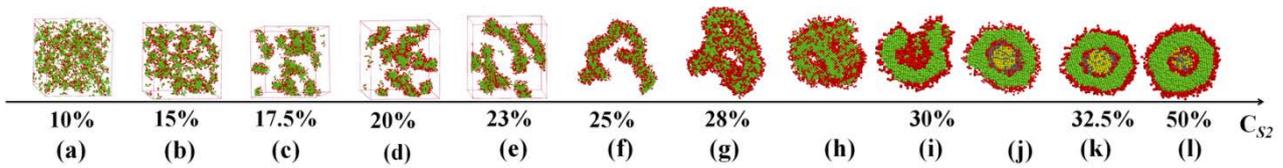

**Figure 6** Morphology sequence of diblock $A_9B_3$ as a function of selective solvent, $C_p = 2.5\%$, $L = 72$ ($\varepsilon_{AS1} = \varepsilon_{BS1} = 0$, $\varepsilon_{AS2} = 3$, and $\varepsilon_{BS2} = 0$). (a) Dissolved, (b) sphere-like, (c) sphere/short-rod, (d) short-rod, (e) rod, (f) rod, (g) ring, (h) cage, (i) open bowl, (j) vesicle, (k) vesicle, and (l) vesicle (green: A monomer; red: B monomer; yellow: common solvent S1; grey: selective solvent S2. The bowl and vesicle profiles correspond to cross sections of cut structures.)

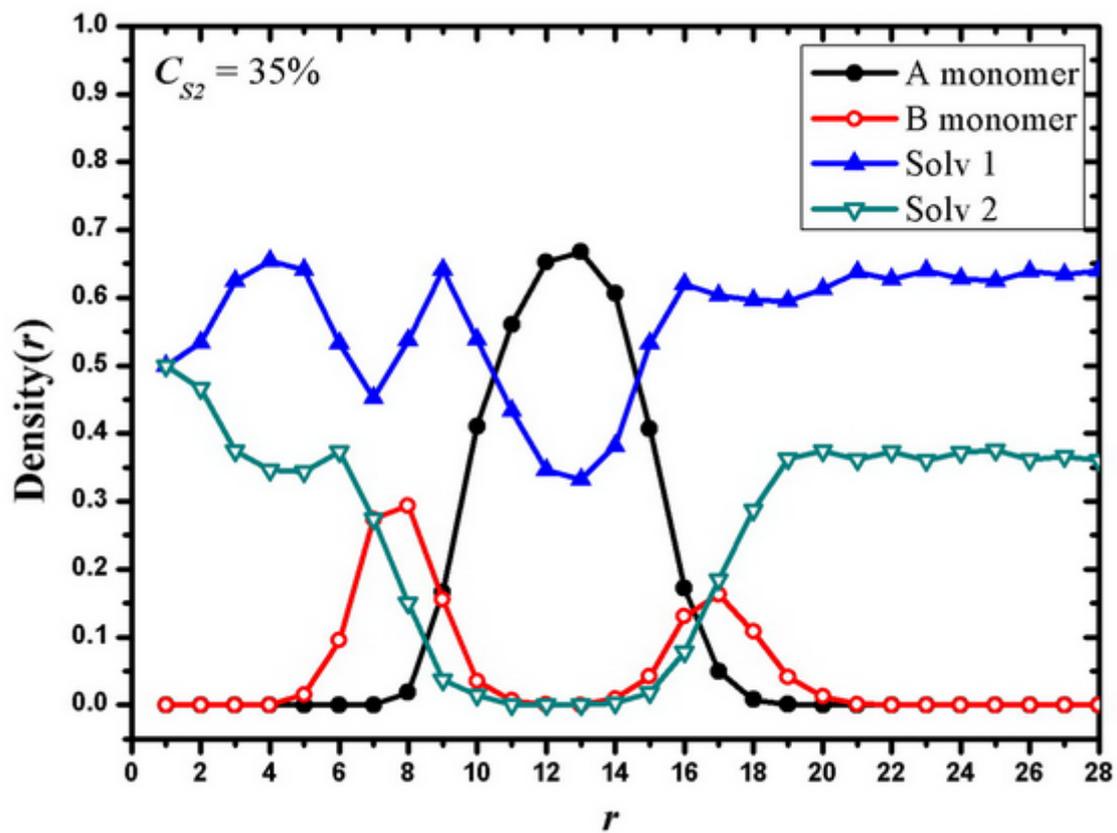

**Figure 7** Variations of the densities of monomers A and B and solvents S1 and S2 with $r$, where $r$ is the distance from the center of mass of the micelle, for a vesicle of $A_9B_3$ at $C_{S2} = 35\%$ ($\varepsilon_{AS1} = \varepsilon_{BS1} = 0$, $\varepsilon_{AS2} = 3$, and $\varepsilon_{BS2} = 0$).

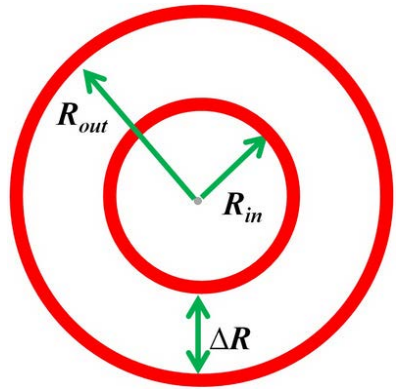

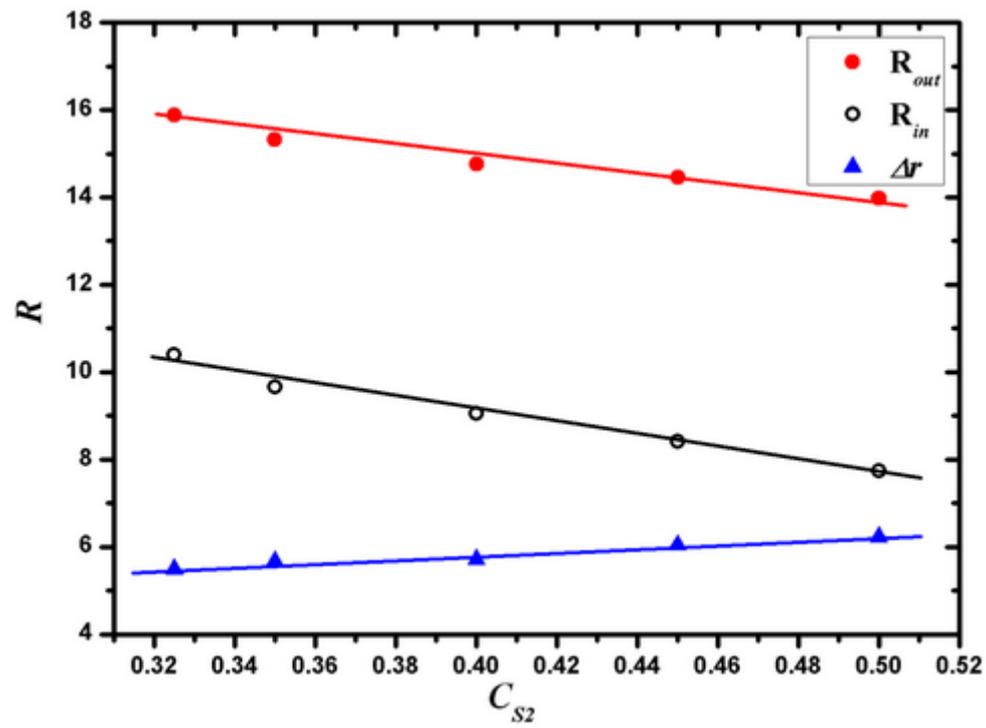

**Figure 8** Size variation of $A_9B_3$ vesicles with increasing $C_{S2}$ for $C_P = 2.5\%$ ($\varepsilon_{AS1} = \varepsilon_{BS1} = 0$, $\varepsilon_{AS2} = 3$, and $\varepsilon_{BS2} = 0$). (a) Schematic of the vesicle structure and (b) radial variation.

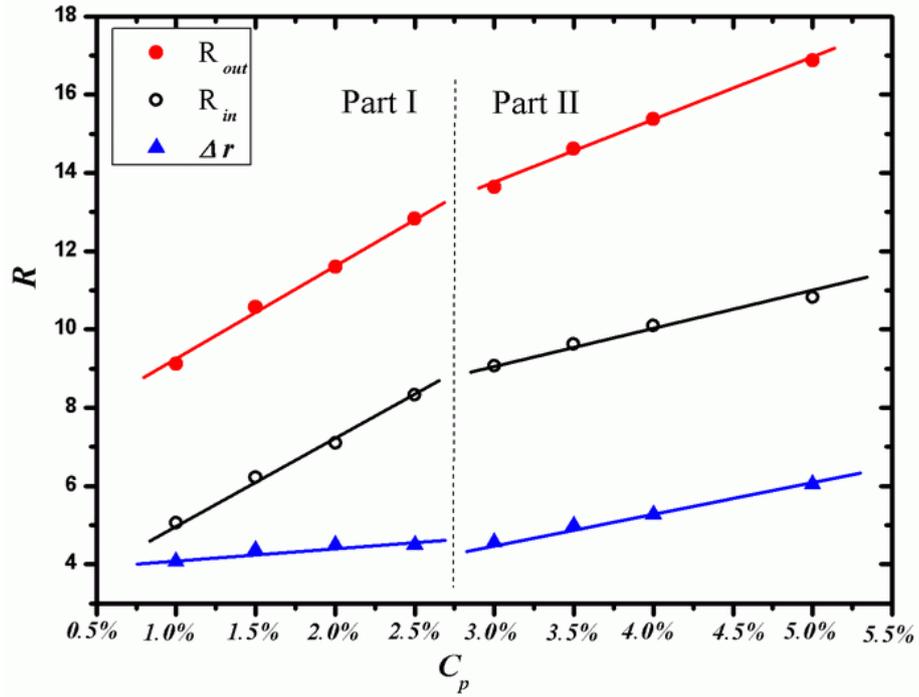

**Figure 9** Variation in radial size of $A_6B_2$ vesicles with increasing $C_p$ for $C_{S2} = 40\%$ ($\varepsilon_{AS1} = \varepsilon_{BS1} = 0$, $\varepsilon_{AS2} = 3$, and $\varepsilon_{BS2} = 0$).

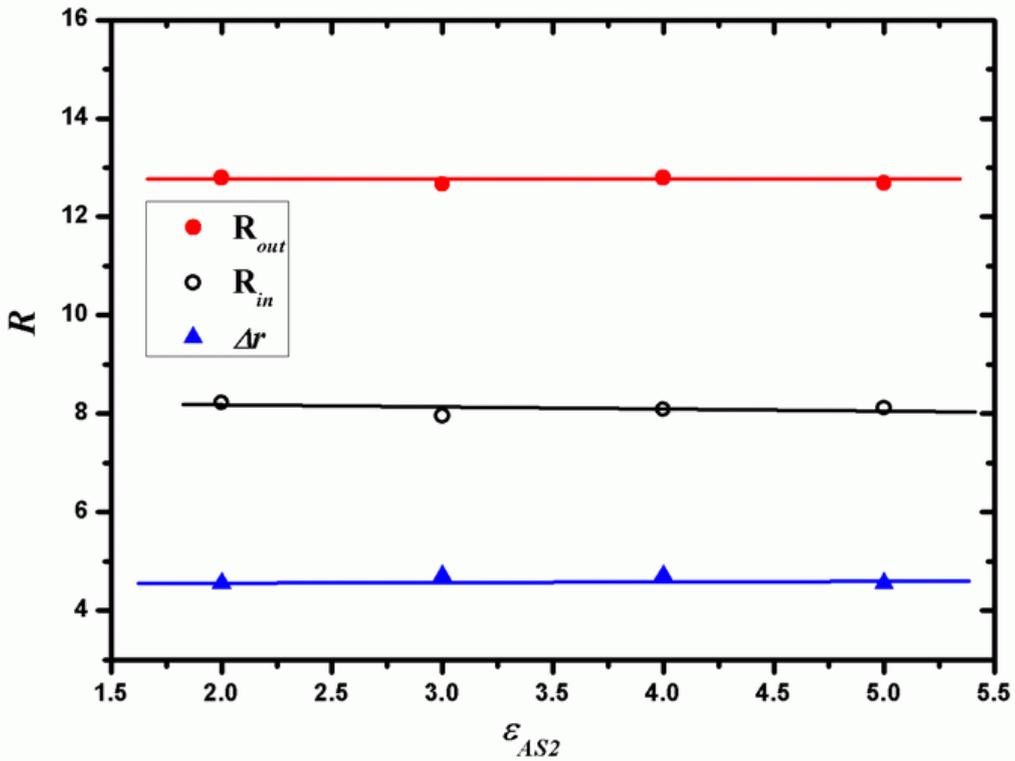

**Figure 10** Variation of radial size of $A_6B_2$ vesicles with increasing repulsive strength $\varepsilon_{AS2}$ for $C_{S2} = 40\%$ and $C_P = 2.5\%$ ($\varepsilon_{AS1} = \varepsilon_{BS1} = 0$, $\varepsilon_{AS2} = 3$, and $\varepsilon_{BS2} = 0$).